\documentclass[12pt]{article}
\usepackage{graphicx}
\usepackage{amsthm}
\usepackage{amsmath}
\usepackage{amssymb}
\usepackage{epsf}
\usepackage{bm}
\usepackage{psfig}
\begin{document}

\title{UNIVERSE DRIVEN BY THE VACUUM OF SCALAR FIELD: VFD MODEL.}
\author{ S.L. Cherkas\thanks{E-mail:cherkas@inp.minsk.by}\\
{\small Institute for Nuclear Problems, Bobruiskaya str. 11,
Minsk, 220050, Belarus}\\
V.L. Kalashnikov\thanks{E-mail:kalashnikov@tuwien.ac.at}\\
{\small Institut f\"{u}r Photonik, Technische Universit\"{a}t
Wien,} \\{ \small Gusshausstrasse 27/387, Vienna A-1040}}
\date{}
\maketitle

\begin{abstract}
It is shown that in the Vacuum Fluctuations Domination model
(VFD), where vacuum fluctuations of scalar fields dominates under
matter and radiation throughout the all history of the  Universe
expansion (gr-qc/0604020, gr-qc/0610148), acceleration parameter
evolves monotonically from the zero to the present day negative
value. That is according to this model the Universe has no
decelerating past and conventional radiation domination and matter
domination epochs are absent. Predictions of accelerating
parameter for $z\sim 0-2$ is compared with that follows from the
SN type Ia data.
\end{abstract}

Fast progress in accumulating and handling of the astrophysical
data about the Universe expansion
\cite{sp99,sp99a,sp99b,wmap,sdss,r} clears the way to testing of
different models of the Universe evolution. Although, the
$\Lambda\mbox{CMD}$ model is able to explain the observational
data \cite{star0}, it is necessary to provide a deeper insight
into the cosmological constant problem
\cite{weinberg,carr,pd,ell,stein,arm,rev,syk}. Among numerous
approaches to the cosmological constant problem, the quantum field
theory (QFT) approach may suggest some solutions.

It is well known that the covariant removing of all divergent
terms from the energy-momentum tensor by some regularization
procedure leads to the vacuum energy density $\rho_{vac}\sim
1/L^4$, where $L$ is the radius of the Universe curvature
\cite{birrel}. This quantity is too small\footnote{For the flat
expanding Universe and the self-interacting scalar field
$V(\phi)\sim\lambda \phi^4$, it is $\rho_{vac}\sim\lambda H^4$
\cite{one}, where $H$ is the Hubble constant. This quantity is too
tiny even for $\lambda \sim 1$.}
 to explain the observed
Universe acceleration if one may identify $L$ with the size of a
present day Universe.

On the other hand, the direct ultraviolet momentum (UV) cut-off for
evaluation of the vacuum energy  provides the enormous quantity
$\rho_{vac}\sim M_p^4$, where $M_p$ is the Planck mass.

In our previous works \cite{1,2}, the accelerated expansion of
Universe was attributed to the back-reaction of vacuum
fluctuations of massless scalar fields. It was found, that the use
of UV cut-off at the Planck level in the equation of motion for
the Universe scale factor instead of that in the Friedman equation
allows explaining the observable value of Universe acceleration.
In our approach, the effective density of dark energy is
proportional to the Hubble constant squared $\rho_{vac}\sim H^2
\kappa_{max}^2\sim H^2 M_p^2$,
 as it occurs in the holographic dark energy models \cite{coh,li,brus,pad,odn,marg} (here
$\kappa_{max}$ denotes the UV cut-off of the present day physical
momentums).

Below our previous model is summarized and compared with the SN Ia
data and gamma ray busts data.

Let us write down the system of Friedman-- Lema\^{\i}tre equations
for the Universe scale factor $a$, the density of a matter $\rho$
and the pressure $p$:
\begin{eqnarray}
-\frac{1}{2}M_p^2\left(a^{\prime 2}+{\mathcal K } a^2\right)+\rho
a^4+\frac{1}{6}M_p^2\Lambda a^4=0,\nonumber\\
M_p^2a^{\prime\prime}=-M_p^2\mathcal K a+(\rho-3 p)
a^3+\frac{2}{3}M_p^2a^3\Lambda,\nonumber\\
\rho^\prime+3\frac{a^\prime}{a}(\rho+p)=0,
\end{eqnarray}
where the conformal time $\eta$ implying the metric
$ds^2=a^2(\eta)(d\eta^2+d\sigma^2)$ is used (the reason will be
explained below), $\Lambda$ is the cosmological constant,
$\mathcal K$ is the signature of space-time and the Planck mass
$M_p$ should be read as $M_p=\sqrt{\frac{3}{4\pi G }}$.

The $\Lambda\mbox{CDM}$ model can be obtained by setting $p=0$,
$\mathcal K=0$ and finally is reduced to the single equation
\begin{equation}
a^{\prime\prime}=2\frac{a^{\prime 2}}{a}-\frac{3}{2}a_0 \mathcal
H^2 \Omega_m,
\end{equation}
where $a_0=a(0)$ is the present day scale factor (this moment
corresponds to $\eta=0$ everywhere below), $\mathcal
H=\frac{a^\prime}{a}\bigr|_{\small \eta=0}$ is the conformal
Hubble constant\footnote{$\mathcal H=H_0 a_0$, where $H_0$ is the
present day Hubble constant.} and the constant $\Omega_m$ is
connected with the matter density $\frac{1}{2}\Omega_m
M_p^2\,\mathcal H^2 a_0=\rho \,a^3=\rho_0 a_0^3$.

Coming to the VFD model \cite{1,2} we set $\Lambda=0$, $p=0$,
$\mathcal K=0$ and add a massless scalar field, which is
characterized by the averaged pressure and the density:
\begin{eqnarray}
\rho_\phi=\frac{1}{V}\int_V\left(\frac{\phi^{\prime 2}}{2
a^2}+\frac{(\bm \nabla
\phi)^2}{2 a^2}\right)d^3\bm r,\nonumber\\
p_\phi=\frac{1}{V}\int_V\left(\frac{\phi^{\prime 2}}{2
a^2}-\frac{(\bm \nabla \phi)^2}{6 a^2}\right)d^3\bm r,
\end{eqnarray}
where $V$ is some volume, which will be set to unity hereafter.
The second step is to turn to the quasiclassical picture, where
the scalar field $\hat \phi(\eta,\bm r)$ is quantum. The resulting
master equations for the VFD model are
\begin{eqnarray}
-M_p^2\frac{a^{\prime 2}}{2}+\rho
a^4+\int\left(\frac{a^2<0|{\hat\phi}^{\prime
2}|0>}{2}+\frac{a^2<0|(\bm \nabla{\hat\phi})^2|0>}{2}\right)d^3\bm r=const,\nonumber\\
M_p^2a^{\prime\prime}=\rho a^3-\int \left(a<0|{\hat\phi}^{\prime
2}|0>-a
<0|(\bm \nabla{\hat\phi})^2|0>\right)d^3\bm r,~\label{sys} \\
\hat\phi^{\prime\prime}+2\frac{a^\prime}{a}\hat \phi^\prime-\Delta
\hat\phi=0,~\nonumber
\end{eqnarray}
where $<0|...|0>$ denotes a mean value over the vacuum state of
scalar field. The first equation is the integral of motion for two
last equations. However, it should be noted that it is not the
Friedman equation because the constant on the right hand side is not
zero. The point is that some renormalization is needed to avoid the
cosmological constant problem, i.e. the huge QFT vacuum energy in
the Friedman equation. Instead of determining the renormalization
constant, one can consider two last equation and fix the constant
assigning the initial condition for the equations. It is very
important, that in conformal time a renormalization is not required
for the second equation. The reason is that the equation contains
exact difference of the kinetic and potential energies of the field
oscillations.
 In the
Minkowski space-time this difference is exactly zero by virtue of
the virial theorem for an oscillator, which states that the
kinetic energy is equal to the potential one in the virial
equilibrium. In the expanding Universe this difference is
proportional to the Hubble constant squared.

Scalar field can be decomposed in a complete set of the modes
$\phi(\bm r)=\sum_{\bm k} \phi_{\bm k} e^{i {\bm k}\bm r}$ and
quantization of the modes consists in postulating  \cite{birrel}
\begin{equation}
\hat \phi_{\bm k}=\hat {\mbox{a}}^+_{-\bm k}\chi_{k}^*(\eta)+\hat
{\mbox{a}}_{\bm k} \chi_{k}(\eta),
\end{equation}
where complex functions $\chi_k(\eta)$ satisfy the relations:
\begin{eqnarray}
\chi^{\prime\prime}_k+k^2 \chi_k+2\frac{ a^\prime}{a}{
\chi^\prime}_k=0,\nonumber\\
a^2(\eta)(\chi_k \,{\chi_k^\prime}^*-\chi_k^*\,\chi_k^\prime)=i.
\label{rel}
\end{eqnarray}

The adiabatic approximation

\begin{equation}
\chi_k(\eta)=\frac{\exp\left({-i \int_0^\eta
\sqrt{k^2-\frac{a''(\tau)}{a(\tau)}} \, d\tau}\right)}{\sqrt{2}
a(\eta) \sqrt[4]{k^2-\frac{a''(\eta)}{a(\eta)}}}
\end{equation}
allows calculating the difference of the kinetic and potential
energies of field oscillators up to the second-order terms:
\begin{eqnarray}
\int \left(a<0|{\hat\phi}^{\prime 2}|0>-a <0|(\bm
\nabla{\hat\phi})^2|0>\right)d^3 \bm r=~~~~~~~~~~~~~~~~~~~~~~~~~~~~~~~~~~~~~~\nonumber\\
\sum_{\bm k}a<0|\hat \phi^\prime_{\bm k}\hat\phi^\prime_{-{\bm
k}}|0>- k^2a <0|\hat\phi_{\bm k}\hat\phi_{-{\bm k}}|0>=\sum_{\bm
k} a({\chi_k^\prime}^{*}\chi_k^\prime-k^2\chi_k^* \chi_k)
\approx\nonumber\\ \frac{1}{2
}\left(-\frac{a^{\prime\prime}}{a^2}+\frac{{a^\prime}^2}{a^3}\right)\sum_{\bm
k}\frac{1}{k }+ O({a^\prime}^3)+O(a^\prime
a^{\prime\prime})+O(a^{\prime\prime\prime}),~~~~~~ \label{15}
\end{eqnarray}
where we imply that $a^\prime$ is the first-order quantity,
$a^{\prime\prime}$ is the second-order one,
$a^{\prime\prime\prime}$ is the third-order one and so on.

Using (\ref{15}) in (\ref{sys}) leads to the master equation of
VFD model in the form:
\begin{equation}
M_p^2 a^{\prime\prime}=\frac{1}{2}\Omega_m M_p^2\mathcal H^2
a_0+\frac{1}{2}\left(\frac{a^{\prime\prime}}{a^2}-\frac{{a^{\prime}}^2}{a^3}\right)\sum_{\bm
k }\frac{1}{k}. \label{ek}
\end{equation}

Eq. (\ref{ek}) can be integrated up to the equation\footnote{This
equation can be also deduced from the first of Eq. (\ref{sys}), when
the corresponding normalization constant is chosen. }
\begin{equation}
a^{\prime^2}=a_0^2 \mathcal{H}^2 {\frac{S_0-1-\Omega_m
\left(a/a_0-1\right)}{S_0a_0^2/a^2-1}}, \label{2a}
\end{equation}
where the parameter $S_0$, from the one hand, is determined by the
UV cut-off $\kappa_{max}$ of the physical momentums $\kappa=k/a_0$
\[
S_0=\frac{1}{2 M_p^2a_0^2}\sum_{\bm k}
\frac{1}{k}=\frac{1}{M_p^2a_0^2(2\pi)^3}\int \frac{d^3\bm k}{2
k}=\frac{\kappa_{max}^2}{8 \pi^2 M_p^2}
\]
and, from the other hand, is connected with the present day
deceleration parameter $q_0$ as
$S_0=\frac{2q_0-2+\Omega_m}{2q_0}$. It was shown \cite{1,2} that
the ultraviolet (UV) cut-off of the present day physical momentums
$k/a_0$ in the sum $\sum_{\bm k }\frac{1}{k}$ at the Planck level
$\kappa_{max}=k_{max}/a_0\sim M_p$ can explain the observed value
of Universe acceleration. In principle, the exact value of UV
cut-off has to result from the Planck scale physics.

A validity range of Eqs. (\ref{ek}), (\ref{2a})  is defined by the
next terms in the expansion (\ref{15}). According to Refs.
\cite{1,2}, the next terms contain additional multiplier
$a^\prime/(a k_{max})$ as compared with the main term, where
$k_{max}$ is the UV cut-off $k_{max}/a_0\sim M_p$ \cite{1,2}. Thus
Eqs. (\ref{ek}), (\ref{2a}) are valid if $\frac{a^\prime}{a}\ll M_p
a_0$, or $\dot a \ll M_p a_0$.

Eq. (\ref{2a}) can also be rewritten in a cosmic time
$dt=a\,d\eta$
\begin{equation}
H^2=\left(\frac{\dot
a}{a}\right)^2=H_0^2\frac{(S_0+\Omega_m-1)a_0^4a^{-4}-\Omega_m
a_0^3a^{-3}}{S_0 a_0^2a^{-2}-1}, \label{ekk}
\end{equation}
which gives $a(t)\approx a_0
H_0\sqrt{\frac{S_0+\Omega_m-1}{S_0}}\,t$ in the vicinity of $t=0$
(i.e. in the conformal time $a(\eta)=\mathcal
H\sqrt{\frac{S_0+\Omega_m-1}{S_0}}\exp\left(\mathcal H
\sqrt{\frac{S_0+\Omega_m-1}{S_0}}\,\eta\right) $). During the
further evolution deceleration parameter
\begin{eqnarray}
q(z)=-\frac{\ddot a a}{\dot a^2}=\frac{1+z}{H}\,\frac{d H(z)}{d
z}-1=~~~~~~~~~~~~~~~~~~~~~~~~~~~~~~~~~~~~\nonumber\\
 \frac{q_0 \left(\Omega_m (2
q_0+\Omega_m-2) z^2+2 \left(\Omega_m^2-3 \Omega_m+2\right)
z+(\Omega_m-2)^2\right)}{(\Omega_m+z (z+2) (2 q_0+\Omega_m-2)-2)
(\Omega_m+z (2 q_0 \Omega_m+\Omega_m-2)-2)},
\end{eqnarray}
comes from zero to the present day negative value at small red
shifts $z=a_0/a-1$ as it is shown in Figs. 1,2.

The parameter $\Omega_m$ amounts 0.27 for both models and
$q_0=-0.8$
 for the
VFD model. These values are chosen to fit the curves within a thin
waist of the experimental data channel near z=0.2.

It is interesting that VFD model is highly insensitive to the dark
matter content. We see that two curves corresponding to the
$\Omega_m=0.27$ and $\Omega_m=0.04$ (pure baryonic matter!) almost
coincide.

\begin{figure}[th]
\centerline{\psfig{file=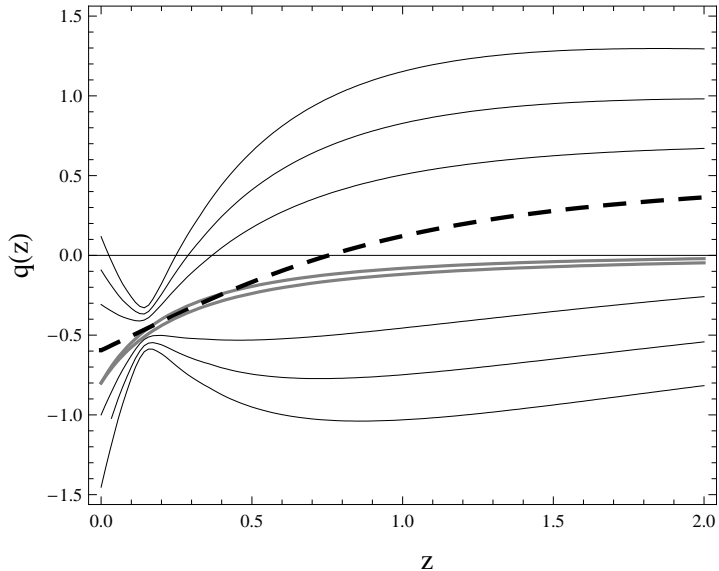,width=9.5cm}} \vspace*{8pt}
\caption{VFD curves (bold grey, $\Omega_m=0.27$ and
$\Omega_m$=0.04) of the acceleration parameter evolution and that
of $\Lambda{CDM}$ (dashed) put on the $1\sigma$, $2\sigma$,
$3\sigma$ error channels (thin lines) of the reconstruction of the
deceleration parameter \cite{kit}  from the 115 SN Ia
data.\protect\label{fig1}}
\end{figure}

It should be noted that in the case of $S_0=0$  our model turns
formally into conventional model of the flat Universe filled with
a dust and a relativistic matter. However, ``matter domination
epoch'' and ``radiation domination epoch'' are in the non physical
region after Big Rip, where the Hubble constant becomes infinite
at some finite $a$ and $t$, when denominator in Eq. (\ref{ekk})
tends to zero.

To summarize, we have considered the VFD model offered in our
previous works \cite{1,2}. In this model, the Universe
acceleration results from the vacuum fluctuations  of fundamental
scalar fields\footnote{According to \cite{2}, there are at least
six fundamental scalar fields including two degrees of freedom of
the tensor gravitational wave.}.

Main feature  of the  VFD model that it does not predict the
change from a deceleration to an acceleration in the past. If the
father observations will insist on such a change, some
modification of VFD should be required, because it has no tuning
parameters. Some possibility of such a modification is a theory
based on the truncation of physical momentums $k/a(\eta)\sim M_p$
rather than that of static momentums $k\sim a_0 M_p$. This would
require a consideration in a system of reference, in which
Universe looks like the Hoyle-Narlikar one \cite{narl,narl1}.

Another feature of the  VFD model is that in principle the dark
matter is not needed.

The authors are grateful to Yungui Gong and Anzhong Wang for kindly
presented the deceleration parameter reconstruction data.

\end{document}